\documentclass[aps,prl,amsmath,amssymb,reprint,superscriptaddress,footinbib]{revtex4-1}

\usepackage{graphicx}% Include figure files
\usepackage{color}

\usepackage{pdfpages}
\usepackage{pgffor}
\makeatletter
\AtBeginDocument{\let\LS@rot\@undefined}
\makeatother

\begin{document}

%\preprint{}

%\title{\kg{Quest for a spin-liquid in CePdAl}}
\title{Entropy evolution in the magnetic phases of partially frustrated CePdAl}
%Magnetic phase diagram of a geometrically frustrated heavy-fermion metal


\author{S. Lucas}
\affiliation{Max-Planck-Institut f\"ur Chemische Physik fester Stoffe, 01187 Dresden, Germany}
\affiliation{Institut f\"ur Festk\"orperphysik, TU Dresden, 01062 Dresden, Germany}
\author{K. Grube}
\affiliation{Institut f\"ur Festk\"orperphysik, Karlsruher Institut f\"ur Technologie, 76131 Karlsruhe, Germany}
\author{C.-L. Huang}
\altaffiliation[Present address: ]{Department of Physics and Astronomy, Rice University, Houston, Texas 77005, United States}
\affiliation{Max-Planck-Institut f\"ur Chemische Physik fester Stoffe, 01187 Dresden, Germany}
\affiliation{Institut f\"ur Festk\"orperphysik, Karlsruher Institut f\"ur Technologie, 76131 Karlsruhe, Germany}
\affiliation{Physikalisches Institut, Karlsruher Institut f\"ur Technologie, 76049 Karlsruhe, Germany}
\author{A. Sakai}
\affiliation{Experimentalphysik VI, Elektronische Korrelationen und Magnetismus, Universit\"{a}t Augsburg, 86159 Augsburg, Germany}
\author{S. Wunderlich}
\affiliation{Max-Planck-Institut f\"ur Chemische Physik fester Stoffe, 01187 Dresden, Germany}
\author{E. L. Green}
\affiliation{Hochfeld-Magnetlabor Dresden (EMFL-HLD), Helmholtz-Zentrum Dresden-Rossendorf, 01314 Dresden, Germany}
\author{J. Wosnitza}
\affiliation{Institut f\"ur Festk\"orperphysik, TU Dresden, 01062 Dresden, Germany}
\affiliation{Hochfeld-Magnetlabor Dresden (EMFL-HLD), Helmholtz-Zentrum Dresden-Rossendorf, 01314 Dresden, Germany}
\author{V. Fritsch}
\affiliation{Experimentalphysik VI, Elektronische Korrelationen und Magnetismus, Universit\"{a}t Augsburg, 86159 Augsburg, Germany}
\author{P. Gegenwart}
\affiliation{Experimentalphysik VI, Elektronische Korrelationen und Magnetismus, Universit\"{a}t Augsburg, 86159 Augsburg, Germany}
\author{O. Stockert}
\affiliation{Max-Planck-Institut f\"ur Chemische Physik fester Stoffe, 01187 Dresden, Germany}
\author{H. v. L\"ohneysen}
\affiliation{Institut f\"ur Festk\"orperphysik, Karlsruher Institut f\"ur Technologie, 76131 Karlsruhe, Germany}
\affiliation{Physikalisches Institut, Karlsruher Institut f\"ur Technologie, 76049 Karlsruhe, Germany}

%\thanks{}
%\affiliation{}


\date{\today}

\begin{abstract}
In the heavy-fermion metal \mbox{CePdAl} long-range antiferromagnetic order coexists with geometric frustration of one third of the Ce moments. At low temperatures the Kondo effect tends to screen the frustrated moments. We use magnetic fields $B$ to suppress the Kondo screening and study the magnetic phase diagram and the evolution of the entropy with $B$ employing thermodynamic probes. We estimate the frustration by introducing a definition of the frustration parameter based on the enhanced entropy, a fundamental feature of frustrated systems. In the field range where the Kondo screening is suppressed the liberated moments tend to maximize the magnetic entropy and strongly enhance the frustration. Based on our experiments, this field range may be a promising candidate to search for a quantum spin liquid. 
\end{abstract}


\maketitle

If competing exchange interactions prevent magnetic systems from developing long-range order, the frustrated magnetic moments can form fluid-like states of matter, so-called spin liquids (SLs) \cite{Balents2010}. If the moments act as effective spin-1/2 particles, quantum fluctuations dominate and impede the moments from freezing or ordering at low temperatures $T$ \cite{Balents2016}. The ground states of these quantum SLs are characterized by massive many-body entanglement rendering them particularly attractive for investigations of new types of quantum matter. Ever since the first notion of SLs was advertised, there has been continual effort to search for materials that might host SLs, mainly in geometrically frustrated magnets \cite{Aharony1976,Chandra1988,Lee2008,Imai2016,Balz2016}. Up to now only very few candidates for metallic SLs have been discovered \cite{Lacroix2010,Balents2016}. 

%\mbox{CePdAl} is a heavy-fermion (HF) metal that displays geometric frustration. 
\mbox{CePdAl} belongs to a class of heavy-fermion (HF) metals with ZrNiAl-type crystal structure (space group $P\bar{6}2m$) that display geometric frustration owing to the fact that the Ce ions form a distorted kagom\'e network in the hexagonal $ab$ plane \cite{Doenni1996,Tokiwa2013,Tokiwa2015}. In HF compounds the magnetic moments are formed by nearly localized 4$f$ or 5$f$ states. 
Magnetic correlations are enabled by the Ruderman-Kittel-Kasuya-Yoshida (RKKY) interaction which competes with the Kondo effect tending to screen the moments at low $T$. The presence of a Kondo effect in \mbox{CePdAl} is manifest through a logarithmic increase of the resistivity with decreasing $T$ \cite{Goto2002,Woitschach2013} and an extremum of the thermopower at low $T$ \cite{Kitazawa1994,Huo2002,Fritsch2016}. 

\begin{figure}
\includegraphics[width = 8.6 cm]{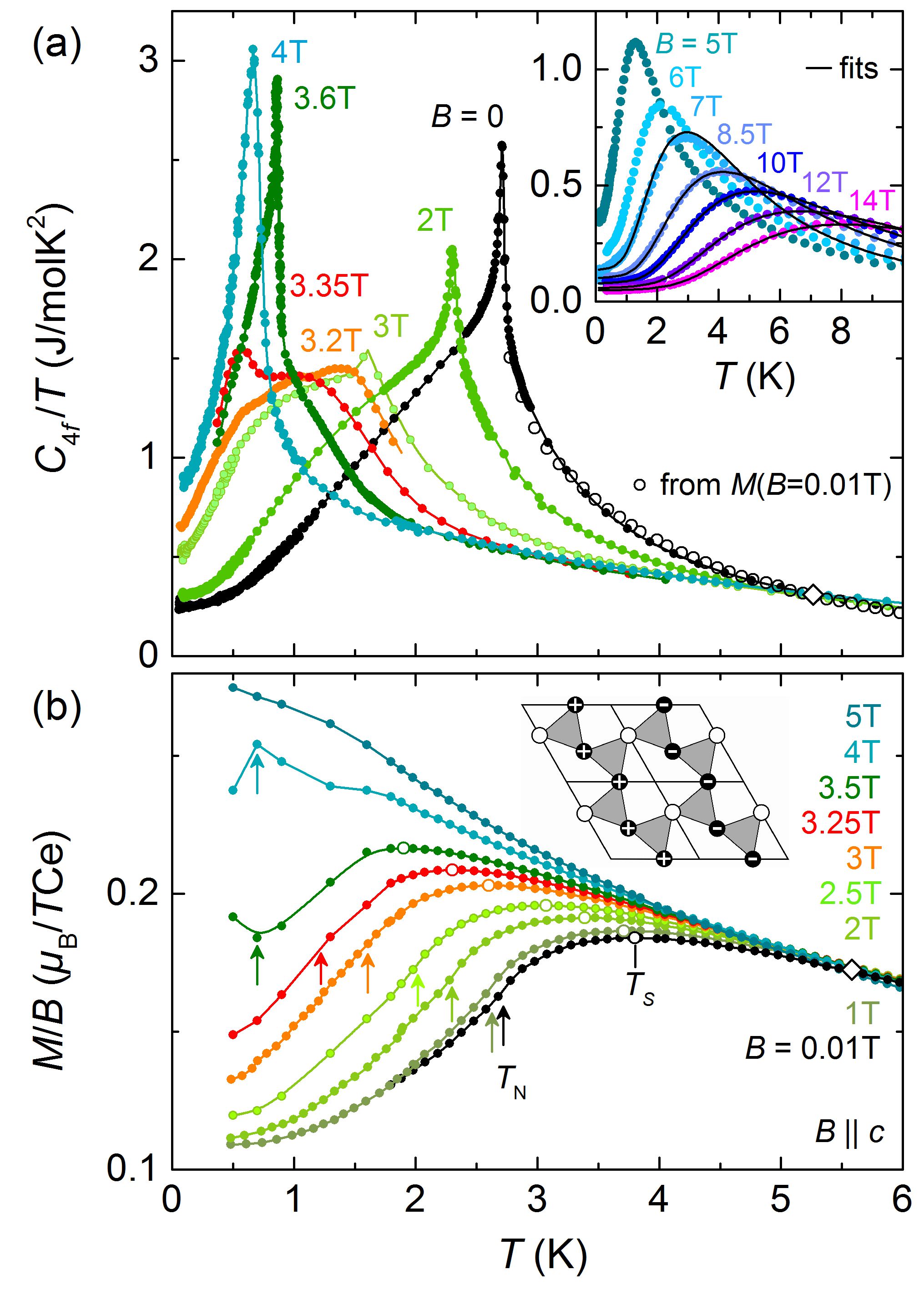}%
\caption{\label{fig:CM} (a) The 4$f$ contribution to the specific heat plotted as $C_{4f}/T$ vs. temperature $T$ for fields along the $c$ axis. The open dots are estimated from the magnetization $M$ at $B=0.01\,$T (see text). The inset shows the Schottky anomaly at higher magnetic fields $B$ with the fits of the resonance-level model for $B\geq 7\,$T (see text). (b) $M/B$ vs. $T$ for different $B$. Open symbols indicate the entropy maximum at $T_S$ and the arrows the magnetic transition determined from $C_{4f}/T$. The magnetic structure of the basal plane is sketched in the inset \cite{Doenni1996,Nunez1997,Fritsch2014}. Crossing points $T^{(C)}_\times$ and $T^{(\chi)}_\times$ of $C_{4f}$ and $M/B$, respectively, are marked by open diamonds. The lines are guides to the eyes.
} 
\end{figure}

\mbox{CePdAl} stands out due to the coexistence of geometric frustration with antiferromagnetic (AF) order below $T_N=2.7\,$K \cite{Kitazawa1994,Doenni1996}. 
Neutron diffraction experiments \cite{Doenni1996} and $^{27}$Al NMR measurements \cite{Oyamada2008} reveal that one third of the Ce moments do not participate in the long-range order down to 30\,mK. Theoretical models considering a quasi-two-dimensional magnetic structure based on the neutron experiments performed on polycrystals \cite{Doenni1996} suggest that the Ce moments of the hexagonal basal plane order in ferromagnetic chains which are antiferromagnetically coupled and separated from each other by the frustrated, interjacent moments [inset of Fig.\,\ref{fig:CM}(b)] \cite{Nunez1997,Fritsch2014}. In the $c$ direction this structure is repeated with an incommensurate AF modulation. Due to the crystal-electric-field (CEF) induced large single-ion magnetic anisotropy between the easy $c$ axis and the basal plane, \mbox{CePdAl} can be regarded as being effectively Ising-like \cite{Isikawa1996}.

As in \mbox{CePdAl} the frustrated moments ($1/3$ of the Ce moments) are correlated but neither freeze nor order and act as effective spin-1/2 particles, this compound fulfills the basic preconditions for a fermionic quantum SL \cite{Lee2008}. It is, however, unclear whether such a state can exist in a HF system, as at low $T$ the Kondo interaction might screen the moments and suppress a possible SL state by quenching the correlations between the frustrated moments without destroying the magnetic order \cite{Lacroix1996,Nunez1997,Oyamada2008,Motome2010,Motome2011}. Even in this case, however, it has been suggested that a SL state may evolve \cite{Senthil2004}. 
Usually, Kondo and RKKY interactions slightly differ in their magnetic-field dependence. Therefore, in an attempt to disentangle geometric frustration from Kondo interaction, we study a Czochalski-grown \mbox{CePdAl} single crystal \cite{Fritsch2016} in magnetic fields up to $B = 14\,$T between $30\,$mK and 10\,K. Here and in the following, $B=\mu_0H$ and $H$ is the magnetic-field strength. As the frustration enhances the degeneracy of the system, we used specific-heat and magnetization measurements to determine the $T$ and $B$ dependence of the entropy $S$. In addition, we tracked the phase boundaries with measurements of the magnetocaloric effect, magnetostriction, and thermal expansion \cite{SM}.
%\footnote{The specific heat, magnetization, and magnetocaloric effect were measured with a Physical and Magnetic Property Measurement System (PPMS) from Quantum Design which has been extended by homemade measurements options. The thermal expansion and magnetostriction was determined by a bespoke capacitive dilatometer built into a dilution refrigerator.}.  

%specific heat
In Fig.\,\ref{fig:CM}(a), we show the $4f$ contribution $C_{4f}$ of the Ce ions to the specific heat $C$ as $C_{4f}/T$ vs. $T$ at several fields. At zero field, $C_{4f}/T$ is dominated by a sharp $\lambda$-like anomaly at $T_N=2.7\,$K. Its asymmetric shape suggests the presence of a shoulder at $\approx 2\,$K that shifts to lower $T$ and becomes more pronounced with increasing $B$. At $B \approx 3.35\,$T, this feature, now being shifted to 0.6\,K, has surmounted that of the by-now broadened transition anomaly at 1.05\,K.
With further enhanced $B$, both peaks vanish and another, even larger transition anomaly appears, see the data for $B=3.6\,$T and 4\,T. Finally, when the magnetic order is suppressed at $B_c(T\rightarrow 0) \approx 4.1\,$T, a Schottky-like anomaly emerges and moves with $B$ to higher $T$ [inset of Fig.\,\ref{fig:CM}(a)]. 
This is due to the field-induced splitting of the Ce$^{3+}$-ground-state doublet. At $B\geq 7\,$T, this anomaly can be described in terms of the single-ion resonance-level model of Zeeman-split quasiparticle levels \cite{Schotte1975}, cf. solid lines in the inset of Fig.\,\ref{fig:CM}(a) \cite{SM}. Recent polarized neutron experiments show that in this field range the intersite correlations are suppressed \cite{Prokes2015}. The temperature of the $C_{4f}/T$ maximum is proportional to the Zeeman energy. A linear extrapolation of its $B$ dependence from fields above $7\,$T not interfered by correlations allows to determine the crossover field $B_K\approx 2.5\,$T. Here, the Kondo screening is supposedly suppressed and, consequently, the two-fold degenerate ground-state levels become energetically distinct. The fact that $B_K$ is well below $B_c$ indicates that the Kondo screening becomes ineffective prior to the suppression of magnetic order. 

%crossing point (1) and magnetization
The geometric frustration becomes apparent by fluctuations visible in $C_{4f}/T$ well above $T_N$. The affected $T$ and $B$ range is marked by a crossing point at $T^{(C)}_{\times} \approx 5.3\,$K and $B < 5\,$T [Fig.\,\ref{fig:CM}(a)] \cite{Vollhardt1997,Eckstein2007}. This point is defined by $\partial (C_{4f}/T)/\partial B=0$ which, according to the Maxwell relation $\partial M/\partial T=\partial S/\partial B$ implies $\partial^2 M/\partial T^2=0$. The related sign change of $\partial^2 M/\partial T^2$ uncovers an increasing deviation from Curie-Weiss-like behavior and the tendency of $M$ to saturate below $T^{(C)}_{\times}$. $M$, indeed, reaches a maximum at a temperature $T_S$ and drops at $T_N$ again [Fig.\,\ref{fig:CM}(b)]. The absence of a Curie-Weiss-like upturn of $M(T)$ at low $T$ demonstrates that the frustrated moments are correlated.

Following the basic approach of Fisher \cite{Fisher1962,Fisher1960}, the strict proportionality between $C_{4f}$ and $\partial [(M/B)T]/\partial T$ for $B\rightarrow 0$, shown in Fig.\,\ref{fig:CM}(a), demonstrates that the extended tails of $C_{4f}/T$ above $T_N$ and the maximum of $M(T)$ are caused by magnetic fluctuations. Notably, as $C_{4f}/T$ does neither saturate nor exhibits a peak at $T_S$, the Kondo effect can be ruled out as source for the maximum of $M(T)$ at $T_S$ \cite{Rajan1983}.

%S(B) maximum
By virtue of the Maxwell relation above, a maximum in $M(T)$ at $T_S$ is equivalent to a maximum of $S(B)$ at $T_S$. An accumulation of entropy naturally emerges when phase boundaries are crossed by employing non-thermal control parameters as the magnetic field \cite{Garst2005}. $T_S$ sensitively depends on critical fluctuations. In frustrated magnetic systems, entropy accumulates at much higher $T$ than the phase-transition temperature \cite{Jongh2001}. In a mean-field description, on the other hand, no difference between $T_S$ and $T_N$ exists. $T_S$ roughly reflects the temperature where the system would order without frustration. Thus, instead of the widely used frustration parameter $f_{CW}=\left|\Theta_{CW}\right|/T_N$ (with $\Theta_{CW}$ the Curie-Weiss temperature) \cite{Ramirez1994}, the ratio $f_S=T_S/T_{N}$ can serve as measure of the frustration strength. In contrast to $f_{CW}$, $f_S$ allows field-dependent studies, as long as $M$ does not saturate at high $B$. Figure \ref{fig:CM}(b) confirms that indeed this is not the case in fields below $5\,$T and the $T$ range of the observed $M/B$ maximum.

\begin{figure}
\includegraphics[width = 8.6 cm]{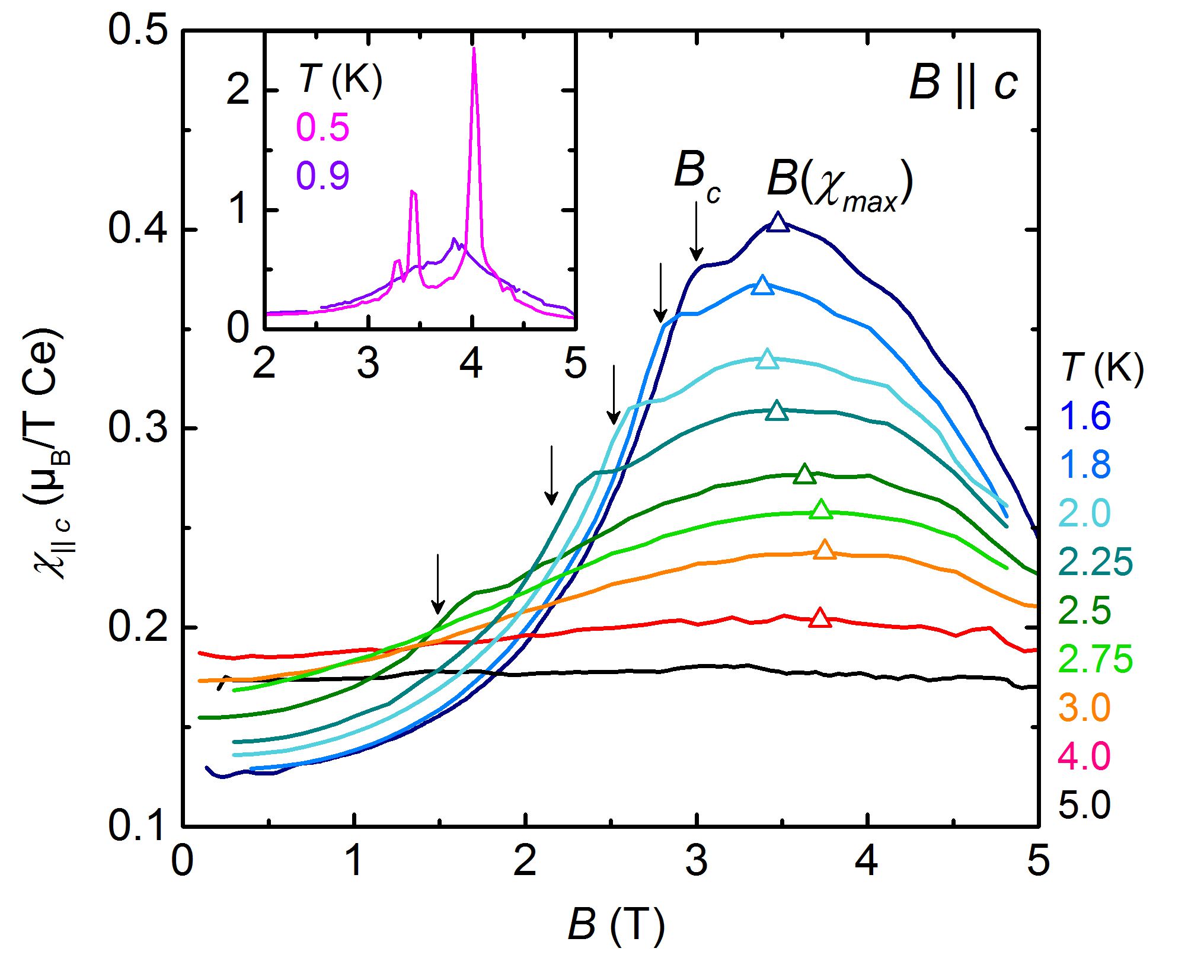}%
\caption{\label{fig:sus} Differential magnetic susceptibility $\chi_{\parallel c}=\partial M/\partial B$ vs. $B$. The inset presents data below $T<1\,$K. The transition fields $B_c$ extracted from $M(T)$ and $C_{4f}/T$ are denoted by arrows and the field of the $\chi_{\parallel c}$ maximum $B(\chi_{max})$ by open triangles.}
\end{figure}

%crossing point (2) and susceptibity
To estimate the field range governed by AF correlations we determined the differential magnetic susceptibility $\chi_{\parallel c}=\partial M/\partial B$ from $M$ measurements for $B \parallel c$ at fixed $T$ (Fig.\,\ref{fig:sus}). 
In accordance with $C_{4f}/T$ and $M(T)$, $\chi_{\parallel c}$ reveals AF correlations above $T_{N}$ as witnessed by the maximum of $\chi_{\parallel c}(B)$ at $B(\chi_{max})$. This maximum is produced by the suppression of the correlations with the magnetic field. 
In the temperature range from $T_{N}$ down to $1\,$K, a single phase transition appears at the critical field $B_c$ as additional shoulder of $\chi_{\parallel c}$. In contrast to usual antiferromagnets, $B(\chi_{max})$ is well separated from $B_c$ and stays roughly constant at $\approx 3.6\,$T (Fig.\,\ref{fig:sus}). Its upper temperature limit is given by the crossing region of $M/B$ at $T_\times^{(\chi)}=5.5\pm 1\,$K and $B<5\,$T [Fig.\,\ref{fig:CM}(b)]. As here $\partial (M/B)/\partial B=0$, $\partial M/\partial B=\chi_{\parallel c}$ is constant.
At further decreased $T<1\,$K, three sharp peaks arise (cf. inset of Fig.\,\ref{fig:sus}) which indicate metamagnetic transitions in agreement with previous measurements \cite{Hane2000}. 

\begin{figure}
\includegraphics[width = 8.6 cm]{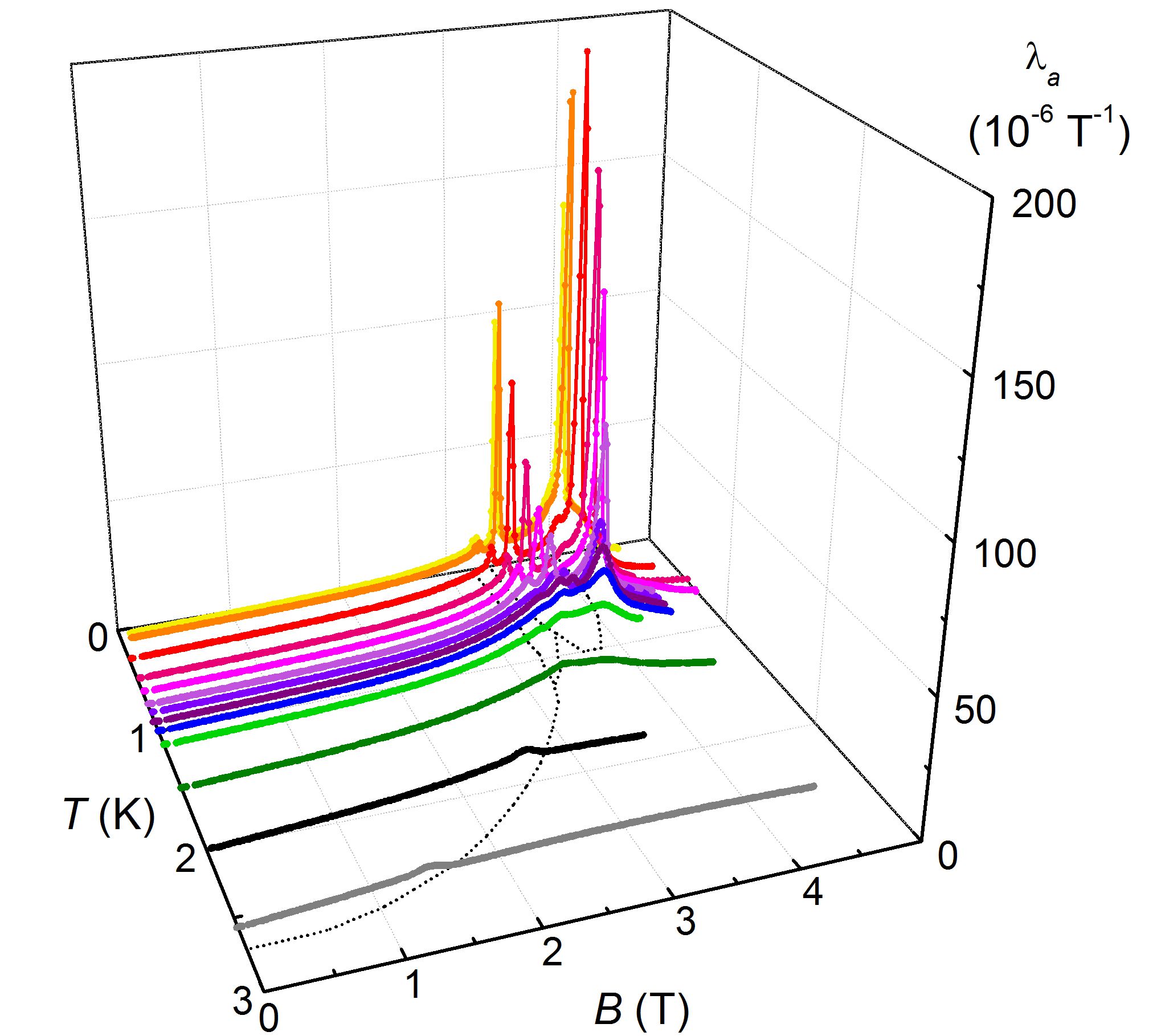} %
\caption{\label{fig:lambda} Magnetostriction $\lambda_a$ of the $a$ axis for fields along $c$. The phase diagram is shown beneath under the data. The discontinuous length changes indicative of first-order transitions give rise to sharp peaks at low temperatures.}
\end{figure}

To establish the phase boundaries we measured the magnetostriction $\lambda_a = (1/L_a)(\partial L_a /\partial B)$ (with $L_a$ as crystal length in $a$ direction) by varying $B$ at fixed $T$ to obtain horizontal cuts through the ($B$,$T$) phase diagram. A three-dimensional plot of $\lambda_a$ vs. $T$ and $B$ is shown in Fig.\,\ref{fig:lambda}. Just as $\chi_{\parallel c}$, $\lambda_a$ clearly reveals three sharp peaks at $T\leq 1$\,K indicative of first-order transitions. At higher $T$, only one transition remains present, visualized by the dotted line in the ($B$,$T$) plane of Fig.\,\ref{fig:lambda}. The change of the height and sharpness of the peaks with increasing $T$ suggests crossovers from first- to second-order transitions.
%The continuous suppression with increasing $T$, together with the decrease of the height and sharpness of the peaks, suggests crossovers from first- to second-order transitions.

\begin{figure}
\includegraphics[width = 8.6 cm]{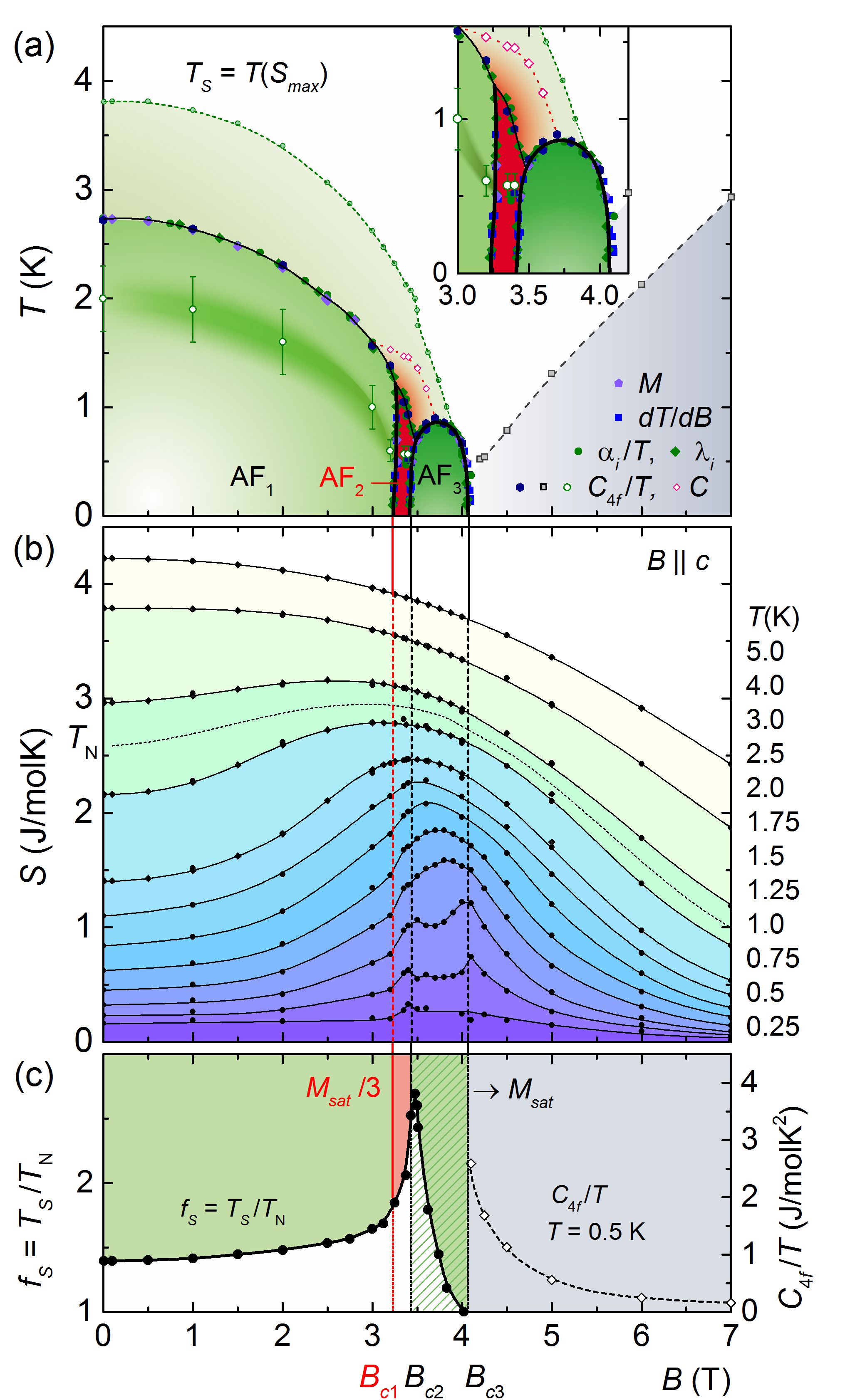}%
\caption{\label{fig:PD}(a) Magnetic phase diagram of \mbox{CePdAl} for $B \parallel c$. Open circles indicate the approximate temperature of the broad shoulders of $C_{4f}/T$ below $T_N$, grey squares the position of the maxima of the Schottky anomalies [Fig.\,\ref{fig:CM}(a)], and red open diamonds the temperature of the $C_{4f}$ maximum. The inset shows an enlarged view on the AF$_2$ and AF$_3$ phases. (b) Entropy $S(B)$ at different, fixed temperatures. (c) Frustration parameter $f_S= T_S/T_N$ vs. $B$. For $B>B_{c3}$, $C_{4f}/T$ at $T=0.5\,$K is plotted against $B$. All lines are guides to the eye.}
\end{figure}

% phase diagram
By extracting the transition temperatures from all data and from additional measurements of the thermal expansion and magnetocaloric effect (not shown), we construct the magnetic phase diagram depicted in Fig.\,\ref{fig:PD}(a), which comprises an extended phase (AF$_1$) between zero field and $B_{c1}= 3.25\,$T, an intermediate phase (AF$_2$) at $B_{c1} < B < B_{c2}=3.4\,\text{T}$, and a smaller pocket (AF$_3$) which ends at $B_c=B_{c3}$. The phase boundaries between AF$_1$ and AF$_2$, and those surrounding the AF$_3$ phase are first-order transitions. The large magnetic anisotropy below $T_N$ (with $10 \chi_{\perp c} \approx \chi_{\parallel c}$) prohibits a canted alignment of the moments \cite{Isikawa1996,Buschow2004} and renders the low-$T$ metamagnetic transitions discontinuous.

%Broader/ rounded peak
The AF$_1$ and AF$_2$ phases show the additional, previously mentioned, shoulder of $C_{4f}/T$ below $T_N$ [Fig.\,\ref{fig:CM}(a)]. As displayed in Fig.\,\ref{fig:PD}(a) and its inset, the shoulder (open circles) becomes more pronounced with increasing $B$ within the AF$_1$ order, shifts to lower $T$ until it reaches $B_{c1}$, and remains constant at $T\approx 0.5\,$K in the AF$_2$ phase. This feature finally terminates at $B_{c2}$, the border to the AF$_3$ phase. Due to its broadness it cannot be attributed to the onset of long-range order but rather points to a crossover. Although the origin of this shoulder is unknown, its field dependence implies a change in the correlations of the frustrated magnetic moments. This conjecture is supported by neutron-scattering experiments which reveal a lock-in of the magnetic propagation vector in a similar temperature range at zero field \cite{Keller2002,Nishiyama2003}. 

% S(B,T) at low T
We note that our data display no signs of additional transitions below the AF phase boundaries. Classical SLs can, however, leave the thermodynamic equilibrium by freezing into disordered spin configurations which are manifested in a finite zero-point entropy \cite{Balents2010,Ramirez1999}. 
We, therefore, determined $S$ as a function of $T$ and $B$ by combining $C_{4f}/T$ and $M$ measurements. The integration of $C_{4f}/T=\partial S/\partial T$ allows to calculate $S(T)$ apart from a field-dependent constant $S_0(B)$. $S_0$ was estimated from $\int \left.\partial M/\partial T\right|_T dB$ using the above Maxwell relation. The remaining integration constant was set to zero at $B>9\,$T where, as mentioned before, the geometric frustration is lifted. 
The derived $S$ values are plotted in Fig.\,\ref{fig:PD}(b). In the entire investigated field range, $S$ approaches zero with decreasing $T$, ruling out a strongly degenerate ground state. 
This opens the possibility that, while below $B_K=2.5\,$T the intersite correlations between the frustrated moments might be removed by Kondo screening, at higher fields a SL dominated by quantum fluctuations develops  \cite{Lee2008}. 
%Although the frustrated moments are present, neither $C_{4f}/T$ nor $\chi_{\parallel c}$ show any divergence but seem to approach constant values at $T\rightarrow 0$, characteristic for Fermi liquids. 
%This again speaks for the evolution of a fermionic SL at fields $B>B_K$ .

% S and \chi maximum
The $S(B,T=\text{const.})$ data clearly reveal a pronounced maximum at $T<3.7\,$K whose position depends on $T$. That temperature is nothing but $T_S$ defined above. With decreasing $T$, it roughly follows the outer AF phase boundaries [Fig.\,\ref{fig:PD}(a) and (b)]. When the AF$_3$ phase is entered below $1$\,K the maximum collapses and two smaller peaks appear that merge into the discontinuous phase boundaries of the AF$_3$ pocket.

% frustration
We are now able to specify the level of frustration by determining $f_S=T_S/T_N$ as a function of $B$ [Fig.\,\ref{fig:PD}(c)], where $T_N$ is defined by the outer phase boundaries. In the AF$_1$ phase, $f_S$ stays almost constant over a wide field range. Upon approaching AF$_2$, however, it starts to grow and shoots up when the phase boundary is crossed at $B_{c1}$. $f_S$ reaches its highest value at the border $B_{c2}$ to the AF$_3$ phase. Beyond $B_{c2}$, the sudden drop of $f_S$ indicates that the frustration is continually removed due to the incipient order of the frustrated moments in agreement with the collapse of $S$ in the AF$_3$ phase. At fields beyond $B(\chi_{max})$ the fluctuations fade out and neither a maximum of $\chi_{\parallel c}$ nor of $S$ can be found. 

According to previous measurements, $M$ exhibits at $T=0.5\,$K three distinct steps as a function of $B$ \cite{Goto2002}, corresponding to the sharp peaks in $\chi_{\parallel c}$ and the even stronger singularities in $\lambda_a$ reported here. At $B_{c1}$, $M$ reaches $1/3$ of the saturated moment $M_{sat} \approx 1.6\,\mu_B$/Ce \cite{Goto2002,Doenni1996}. This suggests that with increasing $B$ the Kondo screening is suppressed at $B_K<B_{c1}$ and that the liberated Ce moments align along the $c$ axis. In a simplified view, the magnetic structure of the basal plane changes from $\uparrow$\,0\,$\downarrow$ to $\uparrow \uparrow \downarrow$. The added, field-polarized moments interfere with the next-nearest-neighbor AF interaction and destabilize the magnetic order at $B_{c1}$. This leads to a significant strengthening of the frustration and an increase of $S$. Here, compared to the other phase boundaries, the transition anomalies are strongly broadened and diminished [Fig.\,\ref{fig:CM}(a)]. The broadening leads to a significant difference between the positions of the anomalies found in $C_{4f}/T$ and $C_{4f}$. The peaks observed in $C_{4f}$ occur at distinctively higher $T$ [red open diamonds in Fig.\,\ref{fig:PD}(a)] and do not coincide with the transition temperatures extracted from $\chi_{\parallel c}$, $\lambda_a$, and thermal expansion measurements. With further enhanced field the frustration increases even more, until the unstable frustrated magnetic structure gives way to the formation of the AF$_3$ phase. This leads to the collapse of $S$ and ultimately lifts the frustration at $B_{c3}$.
%discussion

At still higher fields where $M$ approaches $M_{sat}$, the specific heat still reveals fluctuations, indicated by enhanced $C_{4f}/T$ values at low $T$,  which slowly fade out [Fig.\,\ref{fig:PD}(c)]. These fluctuations, however, originate from the competition between the intersite AF correlations and the ferromagnetic alignment along the applied field as evidenced by the maximum of $\chi_{\parallel c}$. 
%They typically appear in antiferromagnets with a continuous phase transition and thus have to be distinguished from those arising from geometric frustration at lower fields.  

%Outlook and conclusion
In conclusion, our comprehensive measurements show that in \mbox{CePdAl} the geometric frustration persists in a wide field range and is reflected in a rich structure of the entropy $S(B,T)$. Moderate fields gradually suppress the Kondo screening of the magnetic moments. The resulting increase of the frustration and the entropy indicate that the most promising field range to search for a spin liquid is given by the AF$_2$ phase. This phase is characterized by rounded, ill-defined transition anomalies and a prominent shoulder of the specific heat at low temperatures.
To clarify the nature of this phase and the possible existence of a corresponding new type of spin liquid, intertwined with a magnetically ordered solid with competing interactions, further experimental and theoretical efforts are mandatory.

\begin{acknowledgments}
We thank D. A. Zocco, M. Garst, M. Vojta, and R. Eder for valuable discussions. This work was supported by the Deutsche Forschungsgemeinschaft through FOR 960 and SFB 1143, the Helmholtz Association through VI 521, JSPS Postdoctoral Fellowship for Research Abroad, and by the HLD at HZDR, member of the European Magnetic Field Laboratory (EMFL).
\end{acknowledgments}

%\bibliography{CPA_16-12-06}

%

% Include Supplemental Material
\newpage
\foreach \x in {1,2}
 {%
 \clearpage
   

\section*{Supplementary material}

# Supplemental Material to
# Entropy evolution in the magnetic phases of partially frustrated CePdAl

S. Lucas,[1,2] K. Grube,[3] C.-L. Huang,[1,3,4,*] A. Sakai,[5] S. Wunderlich,[1] E. L. Green,[6]
J. Wosnitza,[2,6] V. Fritsch,[5] P. Gegenwart,[5] O. Stockert,[1] and H. v. Löhneysen[3,4]

[1]*Max-Planck-Institut für Chemische Physik fester Stoffe, 01187 Dresden, Germany*
[2]*Institut für Festkörperphysik, TU Dresden, 01062 Dresden, Germany*
[3]*Institut für Festkörperphysik, Karlsruher Institut für Technologie, 76131 Karlsruhe, Germany*
[4]*Physikalisches Institut, Karlsruher Institut für Technologie, 76049 Karlsruhe, Germany*
[5]*Experimentalphysik VI, Elektronische Korrelationen und Magnetismus,
Universität Augsburg, 86159 Augsburg, Germany*
[6]*Hochfeld-Magnetlabor Dresden (EMFL-HLD), Helmholtz-Zentrum Dresden-Rossendorf, 01314 Dresden, Germany*
(Dated: December 8, 2016)

*Experimental details.* The specific heat, magnetization, and magnetocaloric effect were measured with a Physical and Magnetic Property Measurement System (PPMS, MPMS) from Quantum Design which has been extended by homemade measurements options. The thermal expansion and magnetostriction were determined by a bespoke capacitive dilatometer built into a dilution refrigerator. The $4f$ contribution $C_{4f}$ of the Ce ions to the specific heat $C$ was obtained by subtracting $C$ of the non-magnetic sister compound LuPdAl with an empty $4f$ shell and the nuclear contributions of the Pd and Al ions [16].

*Fits of the electronic Schottky anomaly.* An electronic Schottky-like anomaly appears in specific heat data $C_{4f}/T$ at magnetic fields above $B_{c3} = 4.1\,\text{T}$. For Kondo systems, the specific heat in magnetic fields can be described using the single-ion resonance-level model of Zeeman-split quasiparticle levels for a spin-1/2 system [26]:

$$C_{4f}/T = \gamma + k_B \frac{\Delta}{\pi k_B T^2} - \frac{2k_B}{T} Re$$
$$\left\{ \frac{(\Delta+iE)^2}{(2\pi k_B T)^2}\left(4\Psi'(1+\frac{\Delta+iE}{\pi k_B T})\right) - \Psi'(1+\frac{\Delta+iE}{2\pi k_B T})\right\}$$

Thereby, $k_B$ is the Boltzmann constant, $\Psi'$ the derivative of the digamma function, $\gamma$ the Sommerfeld coefficient describing the residual electronic contribution of the conduction electrons, $\Delta = k_B T_K$ the level broadening due to the Kondo effect (Kondo temperature $T_K$) and $E = 2\mu B$ the Zeeman energy of the doublet ground state. Due to the influence of correlations even far above the critical field, reasonable fits are obtained from $B = 7\,\text{T}$ on. The corresponding fit parameters are presented in Table I and agree with the values published recently [16]. Table I also includes the temperature $T_{max}$ of the maximum of the Schottky anomaly in $C_{4f}$ and agrees very well with the relation $k_B T_{max} = 0.42 E$ for a simple two-level Schottky anomaly. From the slope of the Zeeman splitting in the magnetic field, a magnetic moment of $\mu = 1.77\,\mu_B$ is estimated, which is in good agreement with literature data of $1.6\,\mu_B$ [9] and $1.8\,\mu_B$ [27]. Extrapolating the splitting of the Zeeman energy and $T_{max}$ towards $B=0$ finally leads to the crossover field $B_K \approx 2.5\,\text{T}$ above which the Kondo effect is effectively suppressed.

TABLE I. Fit parameters of the electronic Schottky-like anomaly in the magnetic-field range from $B = 7\,\text{T}$ to $14\,\text{T}$. The last row contains the temperature $T_{max}$, where the Schottky anomaly reaches its maximum value in the specific heat $C_{4f}(T)$.

| $B$ (T) | $\gamma\,\left(\frac{\text{mJ}}{\text{mol K}^2}\right)$ | $T_K$ (K) | $E$ (K) | $T_{max}$ (K) |
|---|---|---|---|---|
| 7 | $14.9 \pm 3.2$ | $1.80 \pm 0.08$ | $11.1 \pm 0.1$ | $4.6 \pm 0.2$ |
| 8.5 | $7.4 \pm 1.9$ | $2.53 \pm 0.08$ | $15.0 \pm 0.1$ | $6.1 \pm 0.2$ |
| 10 | $6.2 \pm 1.0$ | $2.83 \pm 0.05$ | $18.3 \pm 0.1$ | $7.7 \pm 0.2$ |
| 12 | $6.1 \pm 0.5$ | $3.23 \pm 0.03$ | $22.9 \pm 0.1$ | $9.5 \pm 0.2$ |
| 14 | $7.9 \pm 0.4$ | $3.54 \pm 0.03$ | $27.4 \pm 0.1$ | $11.7 \pm 0.2$ |


 }

\end{document}